\documentclass{sig-alternate-05-2015}
\usepackage{amsmath}
\usepackage{subfig}

\usepackage{setspace}
\usepackage{lipsum}

\usepackage{graphicx}
\usepackage[T1]{fontenc}

\usepackage{float}

\newtheorem{definition}{Definition}

\usepackage{hyperref}

\makeatletter
\def\@copyrightspace{\relax}
\makeatother
\let\ACMmaketitle=\maketitle
\renewcommand{\maketitle}{\begingroup\let\footnote=\thanks \ACMmaketitle\endgroup}

\begin{document}






%

	\title{Function-Based Access Control (FBAC): From Access Control Matrix to Access Control Tensor\footnote{This paper is a first full version of the paper accepted at ACM CCS 2016, 8th workshop on Managing Insider Threats (MIST). The full version will be submitted at a journal.
		}}
	
\numberofauthors{2} 
%
\author{
%
%
\alignauthor
Yvo Desmedt\\
       \affaddr{University of Texas at Dallas \\ University College London (UCL)}\\
       \email{Yvo.Desmedt@utdallas.edu}
\alignauthor
Arash Shaghaghi\\
       \affaddr{UNSW Australia \\ Data61, CSIRO}\\
       \email{A.Shaghaghi@unsw.edu.au}
}

	
	%


	\maketitle
	
	\begin{abstract}
		
		Security researchers have stated that the core concept
		behind current implementations of access control predates the Internet. These
		assertions are made to pinpoint that there is a foundational gap in this field
		and one should consider revisiting the concepts from the ground up. Insider
		threats, which are an increasing threat vector against organizations, are
		associated to the failure of access control. An in-depth analysis of relevant
		data leakage incidents, such as WikiLeaks, further motivates an outside of the
		box thinking for developing novel and effective countermeasures.
		
		Access control models derived from access control matrix encompass three sets of entities, Subjects, Objects and Operations. Typically, objects are considered to be files and operations are regarded as Read, Write, and Execute. This implies an `open sesame' approach when granting access to data, i.e. once access is granted there is no restriction on command executions. 
		Inspired by Functional Encryption, we propose applying access authorizations
		at a much finer granularity, but instead of an ad-hoc or computationally hard
		cryptographic approach, we postulate a foundational transformation to access
		control. From an abstract viewpoint, we suggest storing access authorizations
		as a three dimensional tensor, which we call Access Control Tensor (ACT). In Function-based Access Control (FBAC),
		applications do not give blind folded execution right and can only invoke
		commands that have been authorized for data segments.
		In other words, one might be authorized to use a certain command
		on one object, while being forbidden to use exactly the same command
		on another object. Obviously, such a behavior can not be modeled using
		the classical access control matrix.

		The theoretical foundations of FBAC are presented along with Policy, Enforcement and Implementation (PEI) requirements of it. A critical analysis of the advantages of deploying FBAC, how it will result in developing a new generation of applications, and compatibility with existing models and systems is also included. Finally, a proof of concept implementation of FBAC is presented.
	\end{abstract}
	
	%
	
	
	\section{Introduction} \label{sec:intro}
	
	We believe there are several reasons as to why we need to fundamentally revise the foundations of access control, and develop models from ground up to overcome existing limitations. The misuse of \textit{legitimate} access to data is a serious information security concern for both organizations and individuals. From a security engineering viewpoint, this is partially due to the failure of access control. To help the reader reflect on the limitations of existing access control, and better understand our motivation for this work, we briefly revise two of the most prevailing cases. \par
	The Wikileaks case was the largest leak of military and diplomatic cables in US history \cite{wiki2011}. After the September 11 terrorist attacks on US soil, government agencies in the United States began allowing a greater sharing of information as a defence procedure against future terrorist strikes \cite{defence}. This included, sharing of confidential information between the US Department of State and the US Department of Defence. However, in 2010 after a massive leak of diplomatic cables by Manning, a low-ranked personnel of the army, the US Department of State revoked this access, to prevent further leaks. As thoroughly explained in \cite{wiki2011}, Manning did not break any system and used his own credentials to access the most sensitive information. Unbelievably, all he had to do was to copy information to a CD drive and take it home. \newline
	As our second example we refer to the case when twenty five Million records of United Kingdom (UK) nationals were lost by an employee of Her Majesty's Revenue \& Customs (HMRC) \cite{schneier2014bruce}. The employee copied the entire available confidential data onto disks and sent it through post \cite{bbc}. \par
	As implied, the main reason behind these incidents is that once user is granted authorization to access data, s/he has the full authority on how to to use it. This is associated to one of today's most prominent security threats, known as Insider Threat \cite{park2006access}. A malicious insider threat is defined as ``when an authorized entity of a system intentionally exceeds or misuses granted access in a manner that negatively affects the confidentiality, integrity, or availability of the organization's information, or information systems'' \cite{cole2005insider}. Recently, insider threats have increased both in number and as a percentage of all cybeattacks; and, various estimates indicate that at least 80 million cases occur in United States per year \cite{harvard}. Evidently, as with the case of access revokation to the US Department of Defence, removing access is not a remedy for this type of security threat. Neither is requiring high security clearance for every officer/employee or enforcing strict limitation, as all of these prevent an organization performing its usual tasks. In this dilemma, an organization to continue its operations has to put trust on its users and this eventually leads to `over-privileged' users phenomena \cite{park2006access, crampton2010towards}. We argue that rather than an `open sesame' approach in access control \cite{desmedt1993computer}, we need models and mechanisms that allow \textit{an authorized entity to perform required operations on confidential information but not have full access to it}. As a simple example, a Department of Homeland Security (DHS) agent should be able to search in confidential information but s/he should only see the relevant information and not be able to run any other operation on it, such as copy and print. At the same time, to ensure information flow control, access restrictions should be applied at the lowest possible level, i.e. data block. Indeed, our ideas are not restricted to text and also applies to images and videos. For example, even when viewing an image - we consider this as being a write operation on the device ``screen'' - only relevant parts of an image must be shown with the non-authorized parts blurred. With existing multi-level security and access control models such as Role-Based Access Control, achieving this type of restrictions is very hard, if not impossible.  \par
	At this point, we discuss some motivational examples from a commercial environment to motivate our argument regarding fundamental gaps in access control even further. Increased infringement of copyright is a serious concern for right holders, including businesses and individuals. For text files, there are a number of tools that can detect copyrighted material. As an example, software such as TurnitIn (Turnitin.com) is now commonly used by universities to detect plagiarism \cite{batane2010turning}. Similar tools exist for images and videos, e.g. see Tineye.com. However, according to \cite{price2013sizing}, copyright infringement is still a growing problem and current mechanisms are not deemed to be effective in reducing it. It is obvious that along with detection, prevention mechanisms are also required. For example, whenever a researcher is preparing a manuscript and quotes a part of the text, both text and citation should be copied. \par
	Due to the pervasive use of portable computers, including laptops and smartphones, many organizations allow, and even encourage, Bring Your Own Device (BYOD) for employees \cite{byod}. With this type of organizational policy, ensuring confidentiality and integrity is challenging. According to studies such as \cite{morrow2012byod, thomson2012byod}, this has resulted in security implications for data leakage, data theft and regulatory compliance. We argue that the fact that existing access control mechanisms are too primitive is one of the reasons for these problems. Today, Apple's App Store, is a bigger business than Hollywood and the number of available applications is increasing day per day \cite{appstore}. To run an application, one needs full execution right and once the application has read and write access to a file, then the application can perform any operation and execute any function on it. Hence, once authorization is granted to a confidential document, there is no control on how this access is used. For example, the user can print, email or share it through other applications. \par
	The root causes of many security problems due to outdated access control have probably been best described by researchers such as Desmedt \cite{rajarajan2012security}, Erlingsson \cite{lee2011advances} and Park and Sandhu \cite{park2004ucon}. In brief, Access Control Matrix (ACM) the core concept behind current implementable systems predates the Internet, computer viruses and massive hackings. At that time of conception, computers had limited resources and there were very limited number of applications. Today, however, there are huge number of applications on each platform with a massive number of functionalities. Moreover, the Internet is only one of the means through which information can leave the user's device. This implies that information flow control mechanisms that rely on entropy to quantify information flow are not reliable by themselves as entropy does not measure the value and the importance of data. At the same time, leakage of a single bit of information could result in loss, or a gain, of ``millions of dollars'' --- we refer the interested reader to the deception plan, Operation Quicksilver, of World War II for understanding the implications of the leakage of one single bit \cite{latimer2001deception, levine2011operation}. \par
	Therefore, we believe it is time for revisiting the foundations of access control, one of the oldest information control mechanisms. It is important to design models that are \textit{compatible with existing access control models} and at the same time can ensure confidentiality and integrity of information in a flexible manner. Inspired by Operator Oriented Encryption \cite{desmedt1993computer} and Functional Encryption \cite{boneh2011functional}, we introduce Function-Based Access Control. From a foundation viewpoint we replace the access control matrix with an Access Control Tensor (ACT), which in effect is a generalization of an access control matrix. In FBAC, objects are data blocks and functions are the commands available in applications, such as Copy/Paste and Search. In the policy specifications of FBAC, the commands may be defined as standard --- as we know them today, or restricted. For example, the Copy/Paste commands could be custom defined such that when a researcher quotes a part of the text that has citation, both the text and citation are copied together to the destination. Or, email function could be customized such that when a sensitive part of a document is emailed, the supervisor is always copied. Moreover, in FBAC, protected objects \textit{do not have to be predefined}, and the function can be customized to protect objects that are created on the fly. In Section VI, a number of examples are shown to describe what this means and why this is a major advantage compared to existing solutions. In our proposed access control model, \textit{applications do not have blind folded execution right and can only invoke commands that have been authorized for data segments in respect to subjects}. To the best of our knowledge, FBAC is the only access control model capable of supporting this level of precision.
	FBAC provides a systematic solution to some of the known failures of access control and replaces adhoc solutions deployed by organizations. The rest of this paper is structured as following. We start with a Background section and then present Function-Based Access Control in Section \ref{sec:fbac}. Thereafter, we discuss Policy, Enforcement and Implementation of FBAC, and in Section \ref{sec:pci} we walk through our prototype implementation. The paper concludes with a critical discussion, where we highlight the advantages, challenges and a number of directions for future work.
	
	\section{Background}
	\subsection{Traditional Access Control Models} \label{trad}
	Access control matrix, introduced in 1971 by Lampson \cite{Lampson74}, remains the core concept for a large fraction of the literature on access control \cite{park2004ucon}. The access control matrix specifies individual relationships between entities wishing access, $Subject (S)$, and the system resources they wish to access, $Object (O)$. For each $S$ and $O$ pair an explicit authorized access, $ (P) $ appears in the corresponding entry in a two-dimensional matrix. The authroization values may include \textit{reading, creating, editing, deleting, and executing} and the objects are files and other system resources. Harrison, Ruzzo, and Ullman \cite{harrison1976protection} identified six primitive operations that transit a system state and established Turing completeness of the access matrix, which shows the expressive power of ACM. As discussed in \cite{sandhu1994access}, the direct implementation of access control matrix is not efficient. However, most access control mechanisms in use are based on models, such as Access Control Lists (ACL) and Capabilities [3, 4], which are derived from the ACM \cite{saunders2001role}. Interestingly, researchers have even formally proved that access control models such as Role-Based Access Control (RBAC) are, in fact, built on top of ACM \cite{saunders2001role}.
	\subsection{Modern Access Control Models} \label{B:modern}
	There has been an increasing concern on the limitations of RBAC in current dynamic and distributed computing environment. Mainly, role explosion - where each role requires different sets of permissions and large number of roles have to be defined - and delays caused due to the role engineering, are limiting factors in the further practice of RBAC \cite{kuhn2010adding}. As a result, a number of extensions have been proposed for this model, e.g. \cite{fadhel2015comprehensive, joshi2005generalized, jin2012rabac}. On the other hand, to overcome limitations of traditional access control, alternative application specific models were also proposed such as relationship based access control \cite{fong2011relationship} and task based access control \cite{oh2003task}. However, all of these extensions and models are purpose built solutions and cannot be generalized into a single framework.
	Attribute-Based Access Control (ABAC) is a general model that associates attributes to subjects and objects. In ABAC, with proper usage of attributes it is possible to have ACL for Discretionary Access Control (DAC), security classifications for Mandatory Access Control (MAC) and roles for RBAC. Moreover, it supports integrating a range of new attributes for access control and having a uniform framework, solves many of the shortcomings of core RBAC \cite{jin2014attribute}. An important advantage of ABAC is that access permissions do not have to be pre-assigned to users and can be computed at the time of request. $UCON_{ABC}$ \cite{park2004ucon} is a conceptual model proposed  by Park and Sandhu for ABAC \cite{jin2014attribute}. In this model, \textit{Authorizations} evaluate subject and object attributes for the requested right, \textit{Obligations} are mandatory requirements for a subject and \textit{Conditions} are system-oriented factors. For instance, \textit{security clearance} is an attribute for authorization, \textit{agreement with the terms and conditions} is an obligation and \textit{the current location} is a condition.\looseness=-1
	\subsection{Access Control with Data-block Granularity}
As mentioned in Section 1, information access control may require applying restrictions based on the content and context related to access requests. Hence, there are an increasing number of publications in the literature that aim to apply access control at the level of document content in different scenarios. A vast majority of these proposals are based on the foundational papers published by Bertino et al., which apply content level protection for XML documents \cite{bertino2001securing, bertino2000specifying, bertino2002protection, bertino2002secure, damianitissec}. Specifically, in \cite{bertino2000specifying}, authors proposed content level access control mechanism for XML documents to enable selective access to data available over the Web. The access control model is described using Document Type Definition (DTD) and considers specific operations, mainly browsing and authoring. However, this work does not provide a general methodology and lacks a role-based model. Moreover, in \cite{bertino2002secure}, Bertino et al. proposed a mechanism to define access policies for XML documents based on user profile and structure and content of a document. They also proposed a mechanism to encrypt different portions of a document with different encryption keys and to selectively distribute the keys among the users based on the access policies. They proposed an architecture to distribute the documents and proved that their scheme generates minimum number of keys. \par

	Recently, Biswas et al. \cite{biswas2015content} proposed a content level access control mechanism for Swift storage service for the OpenStack cloud computing platform. Swift stores outsourced data in a container that is associated with an Access Control List (ACL). This ACL controls the access of the object inside the container. The authors proposed a content level access control on swift object that can be combined with the ACL associated with the container to control the user that can access different parts of an object based on the credential of data requester. The authors utilized JavaScript Object Notation (JSON) to represent data stored in the swift object. They proposed a label based access control to label each JSON item and the data user and then define an access policy to determine the user who can perform certain action on a particular JSON item.
	In \cite{biswas2015content}, the authors utilized the concepts of XML data dissemination in handling JSON data. Moreover, they do not discuss how the view of the data is generated based on access control or whether data encryption is used or not. Memory requirement is huge due to the fact that a large number of JSON items are labeled.
	
	\subsection{Digital Right Management}
Digital Right Management, DRM, is one way of protecting \textit{content that is disseminated}. It was recognized as one of the top ten emerging technologies that will change the world \cite{mitreview}. A fundamental advantage of DRM is separating content from the rights. This enables free distribution of content and then enforcing license procurement for usage \cite{subramanya2006digital}.  A robust DRM system requires a trusted client side reference monitor and uses cryptographic schemes to enforce and monitor access restrictions \cite{zhang2009security}. There was a surge in the number of papers on DRM until early 2000, but mainly due to usability problems, easy bypass methods \cite{zhang2009security}, difficulty in achieving mass scale persistent control, consumer privacy issues, lack of standards, and interoperability of formats, the trend reversed \cite{bird2009global}. DRM is mainly regarded as a collection of enabling technologies, such as watermarking, and lacks proper models and security policies \cite{zhang2009security, bird2009global, park2004ucon}. Due to this, access control and DRM rarely go under the same umbrella. $UCON_{ABC}$ is one of the few models that has tried to integrate DRM into access control.
	
	\subsection{Functional Encryption} \label{FE}
	Operator Oriented Encryption \cite{desmedt1993computer} and Functional Encryption \cite{boneh2011functional} argue that the traditional binary approach in decryption needs to change. In such systems, decryption keys may reveal only partial information about the plaintext. For example, when decrypting an image with a cropping key, a cropped version of image is revealed and nothing more \cite{boneh2012functional}. Boneh defines functional encryption as ``where a decryption key enables a user to learn a specific function of the encrypted data and nothing else. In a functional-encryption system, a trusted authority holds a master secret key known only to the authority. When the authority is given the description of some function $f$ as input, it uses its master secret key to generate a derived secret key $sk[f]$ associated with $f$. Now anyone holding $sk[f]$ can compute f(x) from an encryption of any x'' \cite{boneh2011functional}.
	The main challenge for functional encryption is to ``construct a system that supports creation of keys for any function in both public and non-public index settings'' \cite{boneh2012functional}. Also, efficiency of functional encryption is dependent on specific cryptographic constructions. Overall, although promising, functional encryption is still in its infancy and much further practical and theoretical advancement is required to solve associated open problems.
	\section{Function-based Access Control} \label{sec:fbac}
	We start by contrasting how data is considered by the cryptographic community
	versus how it is considered by these working on access control.
	We will then use this to explain the lessons we want to learn and
	how we can apply these to access control.
	
	We first explain the cryptographic idea of secure multiparty
	computation (see
	e.g.~\cite{Yao86,GoldreichMicaliWigderson87,BenOrGoldwasser88}).
	In this concept, a {\it function\/} is computed by different parties.
	{\it Only the output of this function is leaked and nothing more}.
	We illustrate this concept with the following example.
	Alice, an authorized third party, searches for a string of
	data in files stored inside the Department of Defense or inside the Department
	of State. Suppose there is such a file that contains this string. Then secure
	multiparty computation will only reveal {\it its existence without leaking
		whether this string is on the computers of the Department of Defense, or on
		the Department of State, or on both}.
	
	The second concept we survey is the one of ``operator oriented
	encryption''~\cite[p.~164]{Desmedt93SecPar}, now more known as ``functional
	encryption''~\cite{boneh2011functional}. In functional encryption, given an encrypted text of a
	certain plaintext, one can compute from the ciphertext $f(\hbox{Plaintext})$,
	where $f$ 
	is an authorized function, without revealing anything additionally about the
	rest of the plaintext. As an example, using this tool one could ``search''
	whether a certain string is (or not) in encrypted data without decrypting it. Please refer to Section \ref{FE} for a more detailed definition.
	
	This last example is in sharp contrast with how access to data is being
	controlled today. Indeed, a person searching for the word ``terrorist'' in a
	file, must have received read permission for the file and execute permission
	for the program that does the search. Having the read permission to
	the file is an ``Open Sesame'' approach, giving the person unlimited
	read access to the whole file! In our approach the only thing the user
	will learn is whether the file contains the word ``terrorist'' or not.
	We note that a Unix command as {\it grep\/} (which perform a search in files) facilitates output control, a topic which we will include in our model.
	\subsection{A First Definition} \label{fbac:a}
	
	As also mentioned in Section \ref{trad}, the current approach finds its foundations in the 1974 paper by
	Lampson~\cite{Lampson74} and formalized in 1976 by
	Harrison-Ruzzo-Ullman~\cite{harrison1976protection}.  Its main limitation, from
	our perspective, is that it has only two dimensions, being, one dimension
	corresponding with objects and one with subjects. In our definition we will
	use a three dimensional approach and use ``function'' as the third dimension.
	Note that we regard ``function'' as a synonym for ``operation''.
	
	In our definition, an object could correspond, with a file, an XML record,
	as data in a register, etc. Moreover, functions could be at the level of the
	operating system (such as grep), but also an operation inside an application
	(such as search used inside a browser, an editor, an e-mail reader (or Mail
	User Agents), a global position applications).
	
	Before giving our actual definition we note that the number of inputs to a
	function depends on the function. Our definition has to take this into
	account.  Moreover, not all inputs to a function are ``predefined,'' as we now
	explain.  Consider grep. Usually {\it grep\/} operates on a {\it file\/} and a
	pattern is given, e.g., from the terminal. Moreover, grep has several options,
	such as ``quiet,'' which makes grep output a Boolean.  We do {\it not\/} regard
	the ``pattern'' and the options as objects. We will explain later how to deal
	with these non-object inputs.
	
	To deal with the fact that a function can have more than one object as input
	(such as copy/paste) we introduce the following definition.
	
	\begin{definition}\label{def:o-star}
		When $O$ denotes the set of object, we let $O^1=O$ and recursively
		we define $O^j=O^{j-1}\times O$ ($j\geq 2$). Moreover, we let $O^0=\emptyset$.
		We also define $$O^{*}=\bigcup_{j\geq 0} O^j.$$
	\end{definition}
	
	We now define a first version of Access Control Tensor (ACT).
	
	\begin{definition}
		Let $S$ be the set of subjects, $F$ a set of functions, $O^*$
		as defined earlier. The three-dimensional table $A$
		is a mapping from $S\times F\times O^*\rightarrow \{\hbox{False, True,
			N/A}\}$. When $f\in F$ has $n$ objects as input,
		$o\in O^*$ is an $m$-tuple, $s$ is a subject, then $A(s,f,o)=$ N/A
		when $m\neq n$. If $m=n$, and $A(s,f,o)=$ True
		then subject $s$ can execute the function (command) $f$ on object $o$,
		else the subject can not.
		We call $A$ the access tensor.
		We call $(S,F,O,A)$ an {\it elementary function-based access control},
		or E-FBAC.
	\end{definition}
	Evidently, the set $\{\hbox{N/A, False, True}\}$ could be replaced by
	$\{\hbox{N/A, Forbidden, Authorized}\}$.
	
	One could observe that the typical entries to the Access Control Matrix (ACM),
	such as read and write, do not appear in our ACT. The reason for this is that
	our functions that can read cannot write. Moreover, every read only function can
	be regarded as writing to standard output. So, the function, or the input
	parameters of the function, will define that aspect.
	Note that each command inside an app, such as an editor,
	is regarded as a function and falls under above access control.
	\subsection{The Main Definition} \label{fbac:main}
	The elementary function-based access control is too primitive for many
	different reasons. Let us reconsider {\it grep\/} and assume we allow a user
	in Homeland Security to search files in the CIA for the word terrorist. Using
	the grep option ``context=NUM'' and using a very large value for NUM, the user will
	be able to access the complete file, which might not be the purpose.
	Moreover, the user could use grep to search for other keywords (or in general
	patterns) than the word terrorist. We first discuss how we could fit
	such restrictions in E-FBAC.
	
	Consider we define a new command grep\_terrorist\_count=5, which only allows
	the aforementioned user to search in files for the word terrorist and which
	prints 5 lines of context. In other words this command has no other options.
	Then controlling access when using grep\_terrorist\_count=5 can be described
	using the E-FBAC approach. Obviously, in practice we want the user
	to have the flexibility to use options, which we now address.
	
	\begin{definition}\label{def:G-FBAC}
		Let $S$ be the set of subjects, $F$ a set of functions, $O^*$
		as defined in Definition~\ref{def:o-star}. The entries
		to the three-dimensional table $A$
		with dimensions identified by $S$, $F$, and $O^*$
		are of the type ``False'', ``True$[P_{(s,f,o)}]$,'' and N/A.
		When $f\in F$ has $n$ objects as input,
		$o\in O^*$ an $m$-tuple, $s$ a subject, then $A(s,f,o)=N/A$
		when $m\neq n$. When $m=n$, and $A(s,f,o)=\hbox{False}$,
		the subject can not execute the function (command) $f$ on object $o$.
		In the other case $[P_{(s,f,o)}]$ is an option. If the option is specified,
		then the predefined program $P_{(s,f,o)}$ comprises the joint list of options (with their parameter) together with the
		standard input. If $P$ returns True, then the function $f$ with the
		aforementioned list of options and standard input can be executed
		by $s$ on $o$. We call $A$ the access tensor.
		We call $(S,F,O,A)$ a {\it generalized function-based access control},
		or G-FBAC.
	\end{definition}
	
	Obviously, using G-FBAC in practice might make access control very slow.
	We suggest instead to replace $P_{(s,f,o)}$ by a regular expression.
	If the list of options and the standard input satisfies the regular
	expression, $f$ with the restrictions indicated in
	Definition~\ref{def:G-FBAC}, can be executed. We call this approach
	a {\it regular-expression function-based access control},
	or in short RE-FBAC.
	
	Obviously our approach is very different from the one giving
	subject execution right to functions (or operations) and read/write to objects.
	Indeed, whether an operation can be executed or not should be object
	dependent. To emphasize this aspect of our approach,
	we call this {\it the Function-Data Granularity, or the F-D granularity}.
	It allows to specify that a user can only use ``grep'' with very
	restricted options and patterns on outside data, but allowing
	grep in an unrestricted way on his/her own data.
	
	Before we proceed further in this section, let us make some preliminary
	observations.  As is well known, any three dimensional table can be mapped
	into several two dimensional tables. Indeed, for each (subject,object) we could
	specify which functions could be executed, and provide above restrictions
	specified by $P_{(s,f,o)}$. However, anyone familiar with Lampson's approach
	immediately observes that this does not match the Lampson's description and one also looses the deeper insight the third dimension brings.
	
	It is obvious that our discussion on ``grep'' is just an example and that
	similar OS commands or app commands can be restricted using FBAC. We note that
	the classical Attribute-Based Access Control for XML does not allow us to
	achieve our goal. Indeed, XML organizes the document into ``records.'' When
	granting read access to this record, the {\it maximal output\/} a user can see
	is the whole record. When applying FBAC to an XML document or any other type
	of file, the {\it maximal output\/} a user can see, can contain significantly
	less data than the full record. Finally, the power of FBAC in non-textual
	contexts will be illustrated in the proof of implementation (See Section ~\ref{sec:pci}).
	\subsubsection{Further Output Controls} We first note that in certain contexts it still makes sense to define customized versions of classical commands, such as grep. Indeed, a customized command could further restrict the output by blanking out words or sentences containing predefined words such as ``submarine.'' Unix allows the use of ``pipe'' (i.e., |), which from a mathematical viewpoint
	correspond to a composition of functions, e.g., $f$ after $g$. To regulate
	access to the use of pipe, we could regard $f\circ g$ as a new function
	and then control this as above. We now discuss an alternative
	approach.
	
	If we want to allow the use of pipe and want to avoid having to
	deal with specifying all possible combinations of compositions\footnote{Note
		that the number of different functions one can define with a given
		finite domain is finite, but too large to have practical value.}, the following approach, which we illustrate with
	grep, could be used.
	Grep can be executed on files, but also on standard input, the latter
	enabling to use grep on an output of a prior command when using pipe.
	In our approach the latter use of grep corresponds with a case in which
	grep has no predefined object as input. That implies that we can regard
	grep as being two commands, one being grep\_in\_file and grep\_in\_standard.
	The first has one predefined object as input, the second has zero.
	In the latter, the restriction on the standard input will then be
	specified by the option $P$, as defined in Definition~\ref{def:G-FBAC}.
	\subsection{Access Control Tensor (ACT) in Practice} \label{sub:ac}
	
	Storing rights in an access control matrix is often too impractical or would
	slow down enforcement. Several approaches have been used. Some of these are
	policy dependent, such as the Unix concept of having the {\it user\/} (owner),
	{\it group(s)}, and {\it world\/}, when dealing with access control to
	files. From a conceptional viewpoint, this policy corresponds with a
	compressed authorization list per object. We now wonder what the equivalent
	ones are when using an access control tensor.
	
	In the classical approach, an authorization list corresponds to a column in
	the access control matrix. In other words, given an object, we obtain this
	list. Since our approach is 3-dimensional, given solely an object, the rights
	described related to that object are 2-dimensional, and so it can no longer be
	called a list. We therefore call this an {\it authorization matrix}, i.e., for
	a given object(s) $o$ the authorization matrix gives $A(s_i,f_j,o)$, i.e., all
	values $A(s_i,f_j,o)$ for all $i$ and all $j$. Obviously, we can compress this
	matrix by only considering functions for which $A(s_i,f_j,o)$ will be
	different from N/A.
	
	In operating systems, capabilities play an important role. In our setting
	this will be 2-dimensional and we talk about {\it capability matrix}, or
	just {\it capability}. For each fixed subject $s$ we can have a capability
	corresponding to the matrix $A(s,f_i,o_j)$, which contains
	these values for all $i$ and all $j$. Obviously, we can compress this
	matrix by only considering pairs of (functions,objects) for which
	$A(s,f_i,o_j)$ will be different from N/A.
	
	Obviously, we will have a new 2-dimensional control mechanism, which when
	given a particular function will reveal which subject have rights to which
	objects. Since this matrix has the same dimensions than in the classical
	case, we call this matrix an {\it access control matrix}. In other words, for
	each fixed function $f$ we can have an access control matrix corresponding to
	the matrix $A(s_i,f,o_j)$, which contains these values for all $i$ and all
	$j$. Obviously, we can compress this matrix by only considering object(s) for
	which $A(s_i,f,o_j)$ will be different from N/A.
	
	When systems are large, storing above matrices may be impractical.
	Moreover, when we are using a particular application, only the commands
	(functions) that are available for this application are relevant.
	In such circumstances, we will have two inputs, such as
	$(\hbox{subject},\hbox{object})=(s,0)$,
	and want to know the rights to all (or a subset) of functions.
	We call $A(s,f_i,o)$ given the values for all $i$, a {\it function list}.
	Obviously, we can perform the aforementioned N/A compression.
	If we have an application $P$ and we want to restrict the function list to
	the application, we write $A_{|_{P}}(s,f_i,o)$ to indicate that $f_i$ is a
	function available in the application $P$ and speak of
	{\it application restricted function list}. Note that we can regard the
	commands available in a terminal application, as $P=OS$ or
	$P=\hbox{terminal}$.
	
	For security audits it might be useful to find to know who has access to a
	certain object $o$ when using a function or command $f$.  We call such a list
	a {\it subject list\/} and when given $(f,o)$ it gives $A(s_i,f,o)$ for all
	$i$. When we have a distributed system, we could restrict the subjects to
	$T\subseteq S$.  We denote this restriction as $A_{|_{T}}(s_i,f,o)$.
	(We silently assume that $(f,o)$ is a meaningful pair.)
	
	Finally, when given $(s,f)$ we want to know on what objects the subject $s$
	can execute $f$ and with what restrictions. We call the corresponding
	list an {\it object list\/} it gives $A(s,f,o_i)$ for all
	$i$. When we want to restrict the list of objects to $B^*\subseteq O^*$,
	we have $A_{|_{B^*}}(s,f,o_i)$. $B^*$ may correspond to object(s) 
	inside a certain directory, or objects owned by a certain organization, etc.
	
	The above concepts can be used for all our variants of FBAC, i.e.,
	E-FBAC, G-FBAC, RE-FBAC. \\
	
	\textbf{Extensions:} Our definitions trivially allow to extend the \textit{Harrison-Ruzzo-Ullman} \cite{harrison1976protection} approach, see e.g., \cite[pp.~194-199]{Denning82}. Since this is rather straightforward, we leave the details as an exercise. Note that the primitive operations have to take into account
	that we are dealing with a tensor instead of a matrix. \par
	
	Moreover,in our definition we used $S$ for subject instead of $S^*$. Indeed, cryptographers use the concept of Access Structure, in which trust is put in {\it sets\/} of parties. Replacing $S$ by $S^*$ and using \textit{Access Structures} is beyond the scope of this paper, but deserves a proper study when extending FBAC (see Section~\ref{fw}).
	\section{Policy, Enforcement and Implementation} \label{PEI}
	Up until now, we have formally and theoretically presented FBAC. At this point, we have a discussion on Policy, Enforcement and Implementation of FBAC. Sandhu et al. \cite{sandhu2006secure} have proposed the notion of PEI in an attempt to bridge the gap between abstract policies and real implementations. It should be noted that our discussion in this section, and the next, is one way of implementing FBAC and uses Authorization Matrix to implement the ACT. There are alternative ways of implementing FBAC, which may be more efficient and/or secure and/or suitable. We leave this as future work and present some suggestion in the Future Work section.
	\subsection{Policy} \label{PEI:P}
	Bell-LaPadula~\cite{BellLaPadula73} is a famous approach to model a
	confidentiality policy. Using lattices, its limitations are well
	known (see e.g.~\cite{Bishop03} for a survey on the topic).
	Similarly, the Biba~\cite{Biba77} model is considered the dual for integrity.
	We now explain how, for example, lattice based models, such as Bell-LaPadula
	and Biba can be generalized to FBAC. Note that we do not advocate the use of
	these lattice based models, but that we only show how they could be used.
	
	In a lattice based information flow policy we have a set
	$SC$ of security classes and a relation $\preceq$ on $SC$ such
	that $(SC,\preceq)$ is a lattice. In the case of confidentiality,
	given objects $x$ and $y$ and their corresponding security classes
	$\underline{x}$ and $\underline{y}$, information can flow from $x$ to
	$y$ if $\underline{x}\preceq\underline{y}$.
	In Biba's model, when $s$ is a subject and $o$ is an object,
	$s$ can write when $\underline{o}\preceq\underline{s}$, where
	the $\underline{s}$ is the integrity level of $s$ and similarly
	for $\underline{o}$.
	
	We describe our generalization of the above approaches. We have {\it
		function dependent security classes}. In practice we will specify these
	for subsets of functions.  We now explain the advantages of our function
	dependent policy focusing on confidentiality.
	
	In a military environment we could use strict Bell-LaPadula, but now introduce
	new classes for very special customized functions such as
	grep\_terrorist\_count=5 (or variants further restricting the output) and make
	certain that the appropriate employees at homeland security are in a high
	enough security class for the function grep\_terrorist\_count=5.  Note that by
	having these security classes function dependent, we are able to give
	grep\_terrorist (without count restriction) access to files at the CIA to a restricted number of employees at Homeland Security. So, we can regard
	that to the pair $(f,o)$, where $f$ is a function and $o$ is an object,
	corresponds a security class $\underline{(f,o)}$. Information can
	only flow to subject $s$ if $\underline{(f,o)}\preceq\underline{s}$.
	
	Obviously, we can adapt in a similar manner non-lattice based access control
	policies. We note that we do not see a reason to change the Chinese Wall
	policy. It seems to us that complete separation needs to be maintained
	in circumstances where the Chinese Wall policy is used.
	\subsection{Enforcement and Implementation} \label{enforcandeimpl}
	\subsubsection{Atoms and Atomic Documents} \label{atoms}
	One of the main requirements of implementing FBAC is proper storage of data and authorizations. It is possible to enforce FBAC on any file type as long as the content sections (text and media) are uniquely identifiable. For example, we have used XML in our proof of concept implementation (see Section \ref{sec:pci}). \par
	Once the aforementioned requirement is satisfied, it is possible to have, what we call, an Atomic-Document. An Atomic Document, represented with .\textit{ADoc} extension,  is composed of one or more Atoms. \textit{Atoms} are are the smallest segments of a document and are \textit{undividable}. These could be paragraphs in an unstructured document, a sub-tree in a tree structured data, etc. Atoms have an accompanying policy, which \textit{once executed for a subject} returns a \textit{Function List} $(F)$ --- the policy is an Authorization Matrix but when we regard the matrix for one specific subject then it becomes a \textit{Function List}.  
	\par
	In an Atomic-Document, an Atom can be categorized as being: \par
	\textbf{Single or Linked}: An atom is \textit{Linked} if a function executed on it affects one or more other atoms. In other words, having $f(i)$ and $f(j)$ as a function for $Atom(i)$ and $Atom(j)$ respectively, where $i \neq j$, $E(f(x))$ defined as the execution/invocation of $f(x)$ on $Atom(x)$, and ``$\rightarrow$'' means \textit{results in}, a Linked atom can be defined as when: 
	\begin{equation*}
		E(f(i)) \rightarrow E(f(j)).
	\end{equation*}
	An example of Linked Atom and how it could be used is explained in Section \ref{sec:pci}. \par 
	We define $F(i)$ as the set of allowed functions for $Atom(i)$ and $\overline{F(D)}$ as the set of \textit{not} allowed functions for a document $D$. An $Atom(i)$ is an Atom for document $D$ if and only if the following consistency boolean condition holds: \newline
	\begin{equation}
		F(i) \cap \overline{F(D)} = \{ \}.
	\end{equation}
	Document $D$ is .\textit{ADoc} when the above condition holds for all $Atom(i)$ in document $D$. This condition serves to prevent contradictions. \par
	\textbf{Atomic Document with Classification Level}: To implement access control models such as MAC and security models such as Bell-Lapadula, we require assigning a classification level  to each object. Atomic documents supports defining classification labels for Atoms and .\textit{ADoc} files. In this case, having \textit{C(i)} as classification level of \textit{Atom(i)} and $C(D)$ as classification level of document $D$, we update the consistency boolean Condition 1 as: \par
	\begin{equation}
		F(i) \cap \overline{F(D)} = \{ \} \qquad \wedge \qquad C(i) \subset C(D).
	\end{equation}

	\section{Proof of Concept Implementation: The Smacs Editor} \label{sec:pci}
	
	\begin{figure*}[!t]
		\centering
		\begin{minipage}{.6\textwidth}
			\centering
			\includegraphics[width=9cm,height=4cm]{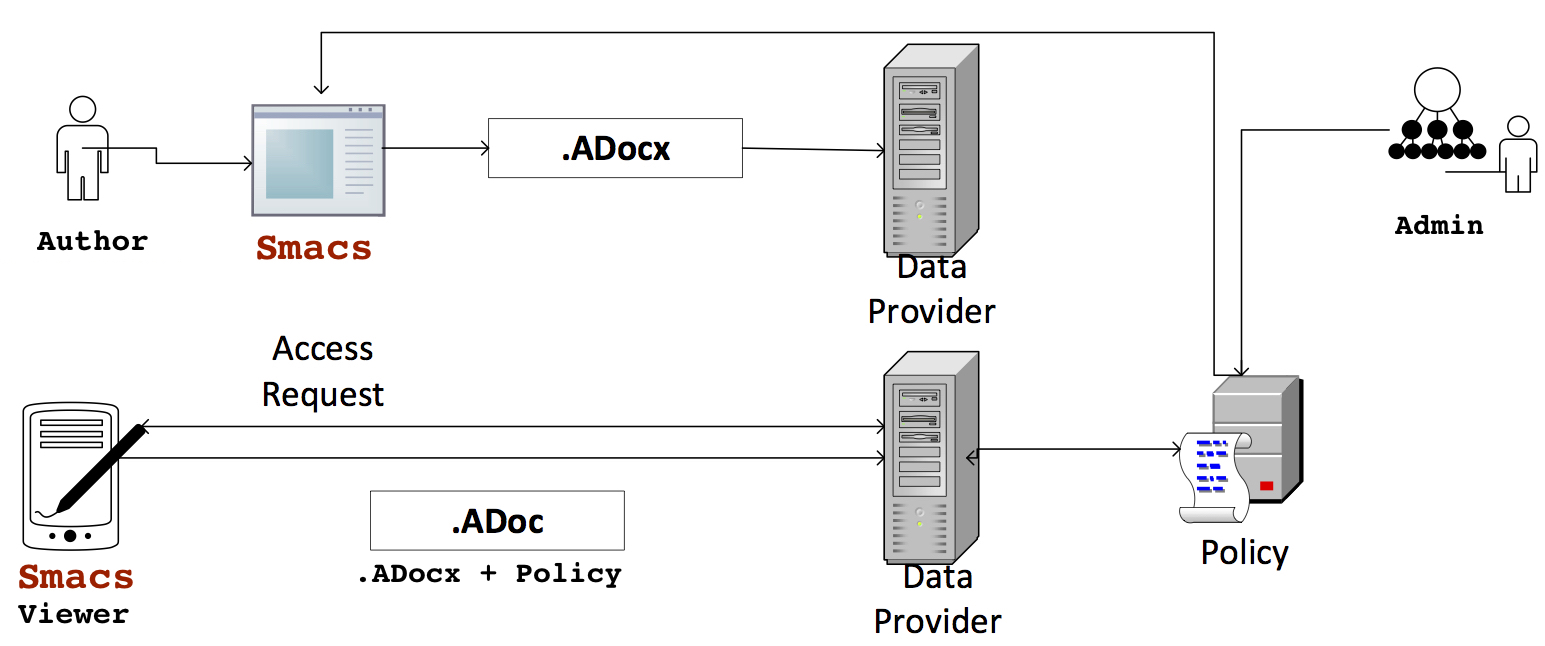}
			\captionof{figure}{One possible Smacs deployment scenario.}
			\label{fig_dep}
		\end{minipage}%
		\begin{minipage}{.4\textwidth}
			\centering
			\includegraphics[width=5cm,height=5cm]{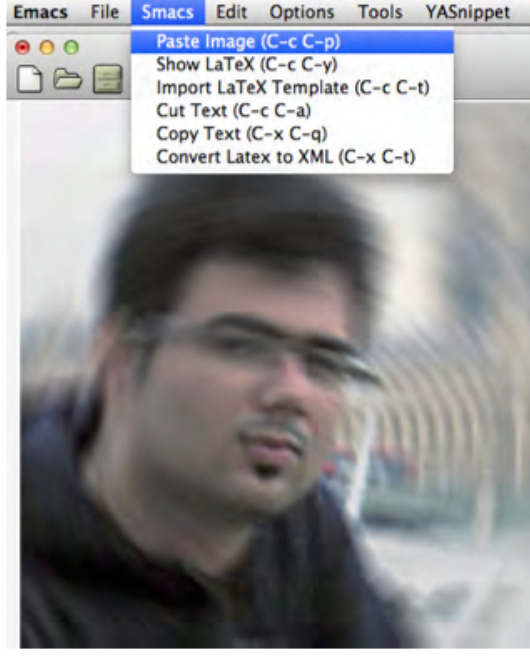}
			\captionof{figure}{Image blurred when relevant atom is not included or the user does not have the right to view it.}
			\label{fig:test2}
		\end{minipage}
		\vspace{0.00mm}
	\end{figure*}
	
	
	As discussed earlier, copyright is a serious concern for right holders. Plagiarism is one of the trending cases related to copyright \cite{guardian, telegraph}, which is considered a case of misconduct in academia and the publishing industry. As suggested by the relevant investigations \cite{savage2004staff, stapleton2012gauging}, limitations of methods for detection calls for innovative preventative mechanisms. We have therefore developed an editor that enforces Function-Based Access Control, \textit{Smacs}, which if used properly could be an effective prevention mechanism against plagiarism. Note that we assume ALL documents are stored in the required format by the editor, see Atomic-Document defined in Section \ref{enforcandeimpl}. We also presume that authors ONLY use the developed editor for creating documents --- an issue we further discuss in Section \ref{fw}. \par
	
	Smacs, or Secure Emacs, is built on top of the GNU Emacs editor. We have created a major mode for Emacs. This mode applies FBAC to both text and images. In Smacs, the access control tensor is implemented as an Authorization Matrix. Once the Atoms, i.e. objects, are defined then the authorization policy is stored as an array in a separate file. Whenever an access request is sent to the reference monitor, in this case Smacs, then the array is processed and retrieves a value, which is a regular expression characterizing the function, as defined in Section \ref{fbac:main}. \par 
	All of the typical commands of Word processing software are available in Smacs. In order to facilitate the user experience when preparing documents, the documents are prepared in \LaTeX format with a custom defined ``\textbackslash\textbf{smacs}'' command. In Smacs, the representation is different from how data is stored and files are saved with .\textit{ADoc} format. The relevant conversion is triggered when accessing and closing the files using an integrated conversion tool. To ensure that only supported functions and commands can be triggered, we had to change the source code of Emacs and recompile it. In this way, we could ensure that no other Emacs mode or commands that could have violated our enforcement mechanism can be executed. In the following, will briefly review sample workflows for three types of users of Smacs, namely the Author, a Co-Author and a Viewer. While doing so, we assume Smacs is deployed in a scenario as depicted in Figure 1. \par
	\textbf{Workflow for an Author:} An \textit{Author} generates a new Atomic-Document, or \textit{ADocx} file, from scratch using Smacs Application. This is then stored at Data Provider servers (see Figure 1). \par
	Currently, for simplicity, the default is set such that each paragraph is regarded as an Atom. However, the author can amend this for any part of the text according to his/her own requirement. An \textit{Author} is asked a set of questions by Smacs so the default \textit{Function List} for Atoms of the Atomic-Document are created. For example, the author is asked whether this document is printable or not. Thereafter, the default \textit{Function List} is assigned a list of \textbackslash Smacs commands throughout the document. The original set of functions that an author can define for atoms may also be specified by an administrator using access control models such as RBAC -- i.e. for each role a set of allowed functions are defined. Indeed, the author at his own discretion, or according to the authority granted by an administrator, may change these. For example, an author may wish to prevent copy on part of the text or require that if this part is printed then his/her name is placed in bold format on top of the page. \par
	To showcase how Smacs works, a number of custom defined functions such as \textit{Watermark-Enforced Print(), Byte-Restricted Copy(), Character-Limited Copy(), Sensitive-Word-Exclusion Copy(), Force-Carbon-Copy Email()} are currently available to an author using Smacs. The author can also specify custom \textit{Search} functions for a document using regular expressions, e.g. \textit{Hide-Sensitive-Word Search()} takes as input a set of words, or Atom unique ID, and hides them from a set of subjects. Or, \textit{Line-Restricted Search()} retrieves a specific number above and below for a query. For a motivational example on the usage of this type of function, see Section \ref{fbac:main}. Evidently, not all of these functions may be required for the plagiarism usecase. \par
	
	\textbf{Workflow for a Co-Author:} A Co-Author is any other user allowed to make changes to the Atomic-Document created by an Author. By default, the set of functions and capabilities available to this user is a subset of those available to the original author. The author, or an administrator, can restrict changing certain parts of the document and could restrict a Co-Author's ability on amending authorized functions. Currently, Smacs supports defining authorization for a global Co-Author and specific policy for each of the possible Co-Authors. \par

	\textbf{Workflow for a Viewer:} A \textit{Viewer} uses Smacs to browse the Atomic-document and retrieves an \textit{ADoc} file stored at Data Provider -- i.e. s/he cannot edit the document. A trusted client-side reference monitor, in this case Smacs, enforces access restrictions for the \textit{Viewer} as per the policies defined by an administrator. Smacs takes as input an \textit{ADoc} file, which contains both the policy and an\textit{ADocX}. First, it computes requirements for the \textit{Read} function for all Atoms. Thereafter, whenever another command available to a viewer such as Print is invoked it refers to the policy file for deciding about authorization. Therefore, if, for example, the Viewer is not authorized to view an image, the image can be hidden, blurred or shown with a watermark -- such features may be useful to prevent unauthorized use of copyrighted images. Figure \ref{fig:test2}, is an example for this case. \par
	Supporting authorization of customized functions and using the Atomic data structure as described earlier, which supports having Linked Atoms (see Section \ref{atoms}) in a document, it is possible to have plagiarism preventive mechanisms. For example, while a Viewer is not allowed to read the document itself, s/he may be allowed to Copy part of the text into another a document that she/he is authoring. \textit{With a customized Copy/Paste function, it is possible to enforce that whenever a text is copied from the document then information is automatically imported as a quote and the source becomes a citation in the destination Atomic document and if the citation is ever removed the quote becomes unavailable.}
	
	\subsection{Usability and Performance Analysis of Smacs} \label{sec:usability}
	The number of features available to each category of users in Smacs, requires careful consideration about the usability aspect. We have customized a number of graphical packages available in Emacs to improve the user experience. When defining authorizations defaults play a major role and, in Smacs, authors can define these by answering a set of questions. When customizing each part of the document a tab is available on the editor window that makes it convenient to change the attributes. Moreover, for any parts that the Viewer is not authorized to read the document information is blacked out and custom error messages are shown when invoking any non-permitted command -- custom messages provide meaningful information and instructions about the error message and minimize disruption of the user experience when using the editor. \par
	In general, the granularity of control provided by FBAC should not be a factor against usability and it should be handled with taste by software developers. Publishing a set of recommendations for applications developed based on FBAC will be done in our following future work. It is also important to note that our performance analysis of current implementation of Smacs compared to the standard Emacs editor, in terms of memory, CPU and responsiveness indicate a negligible performance impact.
	\section{Discussion and Related Work} \label{discussion}
	There is a growing body of literature that takes an incremental approach in detection and prevention of insider threats. These include using monitoring techniques \cite{bowen2009designing}, combining structural anomaly detection with modelling of psychological factors to identify potential insiders \cite{brdiczka2012proactive}, examining behavioural characteristics of potential insiders to distinguish between malicious and benign behaviours \cite{caputo2009detecting} and multi-disciplinary approaches to assist an organisation's analyst in understanding attacks \cite{murphy2012decision, nurse2014critical}.
	Other approaches include using Honeypots to uncover insiders \cite{spitzner2003honeypots}, distributed analysis of data sources, both computer and human factor based \cite{wall2013enemies}, and using using Hidden Markov Models to identify divergence between normal and insider threat patterns \cite{thompson2004weak}. A common argument in this literature is that detection of insider threats is ``not an exact science''. Therefore, we believe these approaches could be regarded as complimentary to our work and that access control is the most critical security mechanism to prevent insider threats \cite{crampton2010towards}. Moreover, with FBAC, due to the level of granularity and practical features of ACT such as \textit{Subject List}, it is possible to narrow down the number of suspects who could have had access to a leaked information much more efficiently. \par
	On the other hand and as mentioned in Section \ref{B:modern}, a number of relevant papers exists in the field of access control. Indeed, models such as $UCON_{ABC}$ have the potential of solving some of the limitations in existing access control. However, to the best of our knowledge, there exist no work until this date that has provided a coherent \textit{model} for authorizing function executions at the level of data blocks. Leave alone, granting custom defined and restricted functions. 
	Cryptographic solutions such as Digital Right Management (DRM) and Functional Encryption that aim to protect content lack proper policy specifications models and standards or are too slow. In addition, we regard DRM and similar software-engineering solutions deployed at organizations to be of an \textit{ad-hoc} approach towards addressing data protection requirements. \cite{zhang2009security} includes some of the main challenges limiting the wider adoption of DRMs. \par
	As mentioned in Section \ref{sec:intro}, entropy does not measure the value of information and we find this literature different in scope with our work. There is also a body of computer security literature such as \cite{zeldovich2006making, myers2001jif, vandebogart2007labels} that provide information flow control solutions at the level of the operating system. This type of work mainly relies on labelling operating system objects and controlling the operating system processes when accessing these objects. These do not consider objects at the level of data blocks and do not target monitoring execution of commands inside applications. Simply put, both the granularity and scope of research is different. However, as we will discuss in the next section when one wants to enforce FBAC at the operating system level, then this literature may become relevant.

	\section{Future Work} \label{fw}
	Our prototype implementation was focused on operations inside one application, being an editor. Emacs was chosen to demonstrate, for example, how the Copy/Paste command could
	dramatically be changed, in particular when writing \LaTeX{} documents.  Several other functions/commands, such as Search, send E-mail, Print, etc., can be used at an OS level, or inside different applications. Hence, if we wish to enforce FBAC properly, there should be no method available on a computer to bypass this. One way to achieve this is to develop an OS where file access controlled by classical access control matrix, is replaced by FBAC. A question worth addressing is to wonder how FBAC can help in practice with controlling information flow inside an OS, i.e., when considering the registers and memory as objects. Having a proper security kernel that facilitates such OS would have several advantages. Given such a security kernel, applications can use the security kernel as a reference monitor to enforce the policy. \par
	Investigating the capabilities of FBAC in addressing selective information sharing requirements in cloud computing and mobile platforms are further directions worth investigating \cite{di2015data}. As a matter of fact, we are currently developing a set of libraries that will allow applications running on Android smartphones to use FBAC and will release this in our future work. \par
	We now discuss what impact our paper may have on the development of new policies. Different policies fit different organizations. However, all current policies are in fact based on a classical access control matrix approach. We have extended some well know policies to adapt them to an FBAC setting. These extensions are rather trivial. Further research may lead to a better understanding how the 3-rd new dimension, i.e., the function, could be exploited to come up with policies to fight insider threats much better, while at the same time allowing flexibility that are currently impossible. Moreover, developing policy specification languages --- such as XACML \cite{moses2005extensible} for attribute-based access models --- properly suited to the requirements of FBAC is an important requirement that has to be addressed in future work. \par
	As stated in Section 3.3, cryptographers interested in secret sharing \cite{Blakley79,Shamir79,ItoSaitoNishizeki88} often regard individuals as untrustworthy, but trust is associated to appropriate subsets of ``parties.'' Due to the Snowden leak, secret sharing is being used for backup purposes. This is a rather limited application.  If one wants to work out this type of approach, a typical subject needs to be replaced by an {\it access structure} \cite{ItoSaitoNishizeki88}, which is a list of subsets. Each subset in this list is trusted.  One of the challenges is on how to implement this. Indeed, let say $\{\hbox{Alice},\hbox{Bob}\}$ are in the access structure.  Does it mean that Alice can only open a file if at exactly the same time Bob tries to do the same. Or should, at the moment, Alice tries to open a file, Bob be notified, and then approve. Such systems have been implemented to enforce very strong audit. However, they have never been formally studied by regarding this as an acces controlled by two parties. More questions arise, such as the fact that access structures contain subsets of parties, and not ordered tuples. If we were to use ordered pairs as $(\hbox{Alice},\hbox{Bob})$ could indicate that Alice is allowed to open a file, provide Bob agree. However, if $(\hbox{Bob},\hbox{Alice})$ is not in the ordered access structure, then Bob might not be able to open the file (i.e., when $\{\hbox{Bob}\}$ is not in the access structure). \par
	Although there has been a lot of progress on functional encryption, that does not mean that there is a cryptographic mechanism to enforce an FBAC policy cryptographically. Such a cryptographic enforcement would correspond with (at least) the use of digital signatures to guarantee that the person {\it granting\/} the rights is authorized.  One of the challenges is to guarantee that when new objects are created from old ones, i.e., combining plaintexts decrypted using functional encryption, access to the new objects will have the correct functional encryption to guarantee the enforcement of the information flow policy, i.e., a re-encryption can not bypass the policy.
	
	\section{Conclusion}

	Mainly motivated by the ongoing insider threats, we changed Access Control Matrix, the core concept behind current implementations of access control, to Access Control Tensor (ACT). We discussed why a 2-Dimensional representation of authorizations is a limitation and argued how our proposed ACT enables achieving a breakthrough level of granularity in access control. We proposed Function-Based Access Control, a new access control model built on top of ACT, which enables designing solutions that could potentially minimize security threats relevant to modern access control failures. In FBAC applications no-longer give blind folded execution rights and access is defined at the level of available commands, such as Copy/Paste, Search, and Email. Commands can be custom defined in FBAC and are applied at the granularity of data segments rather than files.  Finally, we discussed the Policy, Enforcement and Implementation (PEI) aspects of FBAC, provided directions on how to implement, adopt and extend it. 


	
	
	%
	\begingroup
	
	{
	\bibliographystyle{IEEEtran}
	\bibliography{mybib.bib}
	}

	\endgroup

\end{document}